\documentclass[twocolumn,showpacs,amsmath,amssymb,pre,aps]{revtex4}   
\usepackage{graphicx}
\usepackage{dcolumn}
\usepackage{bm}
\usepackage{rotating}

\begin{document}
\title{
Formation and Size-Dependence of Vortex Shells in Mesoscopic Superconducting Niobium Disks 
}
\draft

\author{V.R.~Misko, B.~Xu, and F.M.~Peeters$^{\ast}$}
\affiliation{
Department of Physics, University of Antwerpen, Groenenborgerlaan 171,
B-2020 Antwerpen, Belgium
}

\date{\today}

\begin{abstract}
Recent experiments [I.V.~Grigorieva {\it et al.}, Phys. Rev. Lett. {\bf 96}, 077005 (2006)] 
on visualization of vortices using the 
Bitter decoration technique revealed vortex shells in mesoscopic
superconducting Nb disks containing up to $L=40$ vortices.
Some of the found configurations did not agree with those predicted theoretically. 
We show here that 
this discrepancy can be traced back to the larger disks 
with radii $R \sim 1$ to 2.5$\mu$m, i.e., $R \sim 50-100\xi(0)$ used in the experiment, 
while in previous theoretical studies vortex states 
with vorticity $L \leq 40$ were analyzed for smaller disks with 
$R \sim 5-20\xi(0)$. 
The present analysis is done for thin disks (mesoscopic regime) and for thick 
(macroscopic) disks where the London screening is taken into account. 
We found that the radius of the superconducting disk has a pronounced 
influence on the vortex configuration in contrast to, e.g., the case of 
parabolic confined charged particles. 
The missing vortex configurations and the region of their stability are found, 
which are in agreement with those observed in the experiment. 
\end{abstract}
\pacs{
74.25.Qt, 
74.78.Na 
}

\maketitle

\section{Introduction}

A mesoscopic superconducting disk is the most simple system
to study confined vortex matter where effects of the sample 
boundary plays a crucial role.
At the same time, it is a unique system because just by using disks
of different radii, or by changing the external parameters, i.e.,
the applied magnetic field or temperature, one can cover --- within
the same geometry --- a wide range of very different regimes of
vortex matter in mesoscopic superconductors.

Early studies of vortex matter in mesoscopic disks were focused
on a limiting case of 
thin disks or 
disks with small radii in which vortices arrange themselves in rings 
\cite{GPN97,DSPGL97,lozovik,SPB98,SPL99,BPS01},
in contrast to infinitely extended superconductors where the
triangular Abrikosov vortex lattice is energetically favorable
\cite{tinkham,degennes,abrikosov}.
Several studies were devoted to the questions: 
i) how the vortices are distributed in disks, 
ii) which vortex configuration is energetically most favorable, and 
iii) how the transition between different vortex states occurs. 
Lozovik and Rakoch \cite{lozovik} analyzed the formation and melting 
of two-dimesional microclusters of particles with logarithmic repulsive 
interaction, confined by a parabolic potential. 
The model was applied, in particular, to describe the behaviour 
of vortices in small thin (i.e., with a thickness smaller than 
the coherence length $\xi$) grains of type II superconductor. 
Buzdin and Brison \cite{buzdin} studied vortex structures in
superconducting disks using the image method, where vortices are
considered as point-like ``particles'', i.e., within the London
approximation.
Palacios \cite{palacios} calculated the vortex configurations in
superconducting mesoscopic disks with radius equal to $R=8.0\xi$,
where two vortex shells can become stable.
The demagnetization effects were included approximately by 
introducing an effective magnetic field. 
Geim {\it et al.} \cite{geim} studied experimentally and theoretically
the magnetization of different vortex configurations in superconducting
disks.
They found clear signatures of first- and second-order transitions
between states of the same vorticity.
Schweigert and Peeters \cite{SPL99} analyzed the transitions
between different vortex states of thin mesoscopic superconducting
disks and rings using the nonlinear Ginzburg-Landau (GL) functional.
They showed that such transitions correspond to saddle points in the 
free energy: 
in small disks and rings --- a saddle point between two giant vortex (GV)
states, and in larger systems --- a saddle point between a
multivortex state (MV) and a GV and between two MVs.
The shape and the height of the nucleation barrier was investigated
for different disk and ring configurations.
Milo\v{s}evi\'{c}, Yampolskii, and Peeters \cite{MYPB02} 
studied vortex distributions 
in mesoscopic superconducting disks in an inhomogeneous applied magnetic 
field, created by a magnetic dot placed on top of the disk. 
It was shown \cite{MPB03}, 
that such an inhomogeneous field can lead to the appearance 
of Wigner molecules of vortices and antivortices in the disk. 

In the work of Baelus {\it et al.} \cite{BCPB04} the 
distribution of vortices over different vortex shells in 
mesoscopic superconducting disks was investigated in the 
framework of the nonlinear GL theory and the London theory. 
They found vortex shells and combination of GV and vortex shells 
for different vorticities $L$. 

Very recently, the first direct observation of rings of vortices in
mesoscopic Nb disks was done by Grigorieva {\it et al.}~\cite{grigorieva}
using the Bitter decoration technique.
The formation of concentric shells of vortices was studied for a broad
range of vorticities $L$.
From images obtained for disks of different sizes in a range of
magnetic fields, the authors of Ref.~\cite{grigorieva} traced the
evolution of vortex states and identified stable and metastable
configurations of interacting vortices.
Furthermore, the analysis of shell filling with increasing $L$ allowed
them to identify magic number configurations corresponding to the
appearance of consecutive new shells.
Thus,
it was found that for vorticities up to $L=5$ 
all the vortices are arranged in a single shell. 
Second shell appears at $L=6$ in the form of one vortex in the center
and five in the second shell [state (1,5)],
and the configurations with one vortex in the center remain stable until
$L = 8$ is reached, i.e., (1,7).
The inner shell starts to grow at $L=9$, with the next two states having
2 vortices in the center, (2,7) and (2,8), and so on.
From the results of the experiment \cite{grigorieva}
it is clear that, despite the presence of pinning, vortices generally
form circular configurations as expected for a disk geometry, i.e.,
the effect of the confinement dominates over the pinning. 
Similar shell structures were found earlier in different systems as 
vortices in superfluid He \cite{campbell,hess,stauffer,totsuji,tsuruta}, 
charged particles confined by a parabolic potential \cite{BePB94}, 
dusty plasma \cite{plasma}, 
and colloidal particles confined to a disk \cite{colloids}. 
Note that the behavior of these systems is similar to that of vortices in thin 
mesoscopic disks, thus our approach of Sec.~II can be used for better understanding 
of the behavior of various systems of particles confined by a parabolic potential 
and charachterized by a logarithmic interparticle interaction (e.g., vortices in 
a rotating vessel with superfluid He \cite{campbell,hess,stauffer,totsuji,tsuruta}). 
In contrast, our results presented in Sec.~III are specific for vortices 
in thick large mesoscopic superconducting disks, where the London screening 
is important, and the intervortex interaction force is described by the modified 
Bessel function.

The filling of vortex shells was experimentally analyzed 
\cite{grigorieva}
for vorticities up to $L=40$.
Many configurations found experimentally agree with earlier numerical
simulations for small $L$ which were done for mesoscopic disks with
radii as small as $R \approx 6-8\xi(0)$,
although the disks used in the experiments \cite{grigorieva}
were much larger, $R \approx 50-100\xi(0)$.
At the same time,
some theoretically predicted configurations were not found in the
experiment, such as states (1,8) for $ L = 9$ and (1,9) for $L=10$. 
The difference between vortex states in small and large disks
becomes even more striking for larger vorticities $L$.
In small disks with radii of a few $\xi$, the formation of 
GVs is possible if the vorticity $L$ is large enough, e.g., 
in disks with $R=6\xi$, a GV with $L=2$ appears in the center 
for total vorticity $L=14$, but for vorticities $L>14$, 
{\it all} the vortices form a GV \cite{BCPB04}. 
Obviously, this boundary-induced formation of GVs is possible 
only in the case of small disks: in large disks vortices instead 
form the usual Abrikosov lattice which is distorted near boundaries. 

The aim of the present paper is to theoretically analyze vortex states, 
using Molecular-Dynamics (MD) simulations, 
in rather {\it large} mesoscopic superconducting disks 
and thus to study the crossover between mesoscopic and macroscopic disks, 
i.e., the regime corresponding to the Nb disks used in 
the recent experiments of Ref.~\cite{grigorieva} 
and to look for the missing vortex configurations in the
earlier simulations. 
We analyzed the region of stability of those configurations and 
performed a systematic study of all possible vortex configurations. 
We found that the radius of the disk has an influence on the vortex 
shell structure, 
in contrast to the case of charged particles confined by a 
parabolic potential \cite{BePB94}. 
This analysis was done for thin mesoscopic disks and for thick disks 
where the London screening becomes pronounced. 
We also perform calculations of vortex configurations using the 
GL equations, and we compare these results to the ones 
obtained within the MD simulations. 
The calculated vortex configurations agree with those observed 
in the experiment \cite{grigorieva}. 

The paper is organized as follows.
In Sec.~II, we discuss thin mesoscopic disks. 
The model and the simulation method is described in Sec.~II.A. 
In Sec.~II.B, we discuss different vortex configurations and the formation
of vortex shells.
We analyze the ground states and metastable states for different
vorticities in Sec.~II.C, using a statistical study, similar to the one 
employed in the experiments of Ref.~\cite{grigorieva}, by starting with
many random vortex configurations
and comparing the energies of different vortex configurations.
Based on that analysis we reconstructed the ``radius $R-$magnetic field $H$''
phase boundary (Sec.~II.D).
In Sec.~III, we study the crossover from thin mesoscopic to thick macroscopic 
disks, and we analyze the impact of the London screening on the vortex 
patterns. 
In Sec.~IV, we calculate the crucial vortex configurations in disks using
the GL equations, and we compare them to the results obtained within
the MD simulations.
A summary of the results obtained in this work is given in Sec.~V.

\section{Mesoscopic disks}

\subsection{Theory and simulation}


In this Section, 
we consider a thin disk with thickness $d$ and radius $R$ 
such that $\lambda_{eff} \gg R \gg \xi \gg d$, 
placed in a perpendicular external magnetic field $\bm{H}_0$.
Here $\lambda_{eff} = \lambda^{2}/d$ is the effective London
penetration depth for a thin film, $\lambda $ is the bulk London
penetration depth, and $\xi$ is the coherence length. 
We follow here the theoretical approach developed in 
\cite{CBPB04,BCPB04,buzdin} for thin disks and we use the 
original dimensionless variables used in those works. 
Thus, following \cite{CBPB04,BCPB04}
the lengths are measured in units of the coherence length $\xi$, 
the magnetic field in units of 
$H_{c2} = c \hbar / 2e \xi^{2} = \kappa \sqrt{2} H_{c}$, 
and the energy density in units of $H_{c}^{2}/8\pi$. 
The number of vortices, or vorticity, will be denoted by $L$.
In a thin disk in which demagnetization effects can be neglected 
the free energy in the London limit can be expressed as
\cite{CBPB04,BCPB04,buzdin}
\begin{equation}
G_{L}=\sum_{i=1}^{L} \left(
{\epsilon}_{i}^{s}
+\sum_{j=1}^{i-1}{\epsilon}_{ij}\right)+{\epsilon}^{core}
+{\epsilon}^{field},
\label{fel}
\end{equation}
where the potential energy of vortex confinement 
consists of two terms: 
\begin{equation}
{\epsilon}_{i}^{s}={\epsilon}_{i}^{self}+{\epsilon}_{i}^{shield}
\label{epss}
\end{equation}
i.e., the interaction energy between the $i$th vortex and the 
radial boundary of the superconductor, 
\begin{equation}
{\epsilon}_{i}^{self}=\left(\frac{2}{R}\right)^{2}ln (1-r_{i}^{2}), 
\label{epsself}
\end{equation}
and 
the interaction energy between the $i$th vortex and the
shielding currents, 
\begin{equation}
{\epsilon}_{i}^{shield}= -2 H_{0}(1-r_{i}^{2}). 
\label{epsshield}
\end{equation}
In Eq.~(\ref{fel}), 
\begin{equation}
{\epsilon}_{ij}= \left(
\frac{2}{R}\right)^{2}ln \left[ \frac{(r_{i}r_{j})^{2}
-2{\rm \bf r}_{i}\cdot {\rm \bf r}_{j}+1}
{r_{i}^{2}-2{\rm \bf r}_{i}\cdot
{\rm \bf r}_{j}+r_{j}^{2}}\right]
\label{epsij}
\end{equation}
is the repulsive interaction energy between vortices $i$ and $j$.
Here
${\rm \bf r}_{i} = {\rho}_{i}/R$
is the distance to the vortex normalized to the disk radius. 
The divergence arising when $i = j$ is removed in Eq.~(\ref{epsij}) 
using a cutoff procedure 
(see, e.g., \cite{abrikosov,book,buzdin}) 
which assumes the replacement of 
$ \mid {\rho}_{i} - {\rho}_{j} \mid $ 
by 
$a$ 
(or by $a \xi$ in not normalized units) 
for $i = j$. 
Finally,
$\epsilon^{core}=(2/R)^{2}L$ $ln(R/a)$
and
$\epsilon^{field}=R^{2}H_{0}^{2}/4$ 
are the energies associated with the vortex cores 
and the external magnetic field, respectively. 
Notice that the energy of the vortex cores $\epsilon^{core}$ 
becomes finite due to the cutoff procedure and is strongly dependent 
on the cutoff value $a \xi$ \cite{BCPB04}. 
Here we use for the vortex size $a=\sqrt{2}\xi$; this choice, as shown 
in Ref.~\cite{CBPB04}, 
makes the London and the Ginzburg-Landau free energies to agree 
with each other. 

From the expression of the free energy given by
Eqs.~(\ref{fel}-\ref{epsij}),
we obtain the force acting on each vortex,
$ -\nabla_{k} G(\boldsymbol{\rho}_{i},\boldsymbol{\rho}_{j}),$
where
$ - \nabla_{k}$
is the gradient with respect to the coordinate
$\boldsymbol {\rho}_{k}$.
This yields a force per unit length,
\begin{equation}
{\rm \bf F}_{i} = {\rm \bf F}_{i}^{s} +
\sum_{k \neq i}{\rm \bf F}_{i,k}^{int},
\label{fi}
\end{equation}
in units of
$H_{c}^{2}\xi/8\pi$,
where the summation over $k$ runs from 1 to $L$, except for $ k = i$.
The first term describes the vortex interaction with the current
induced by the external field and with the interface,
\begin{equation}
{\rm \bf F}_{i}^{s}=\left( \frac{2}{R}\right)^{3}
\left(
\frac{1}{1-r_{i}^{2}}-\frac{H_{0}R^{2}}{2}\right){\rm \bf r}_{i}.
\label{fis}
\end{equation}
The second term is the vortex-vortex interaction
\begin{equation}
{\rm \bf F}_{i,k}^{int}=\left( \frac{2}{R}\right)^{3}
\left(\frac{{\rm \bf r}_{i}-{\rm \bf r}_{k}}{ \mid {\rm \bf r}_{i}-{\rm \bf r}_{k}|^{2}}
-r_{k}^{2}\frac{r_{k}^{2}{\rm \bf r}_{i}-{\rm \bf r}_{k}}
{\mid r_{k}^{2}{\rm \bf r}_{i}-{\rm \bf r}_{k}|^{2}}\right).
\label{fivv}
\end{equation}
The above equations allow us to treat the vortices as point-like 
particles and the forces resemble those of a two-dimensional system 
of charged particles with $1/r$ repulsive interaction confined by 
some (usually, parabolic) potential \cite{BePB94}. 
However, the inter-vortex interaction in our system is different 
from $1/r$, and the confined potential differs from parabolic 
and depends on the applied magnetic field. 

To investigate different vortex configurations, we perform
MD simulations of interacting vortices in a disk
(see, e.g., Ref.~\cite{CBPB04}), 
starting from randomly distributed vortex positions. 
The final configurations were found after typically $10^{6}$ MD steps. 

In order to find the ground state (or a state with the energy very 
close to it) we perform many (typically, one hundred) runs of simulations 
for the same number $L$ of vortices starting each time from a different 
random initial distribution of vortices. 
As a result, we obtain a set of final configurations which we analyze 
statistically, i.e., we count probabilities to find the different 
configurations with the same vorticity $L$, e.g., configurations (1,8) 
and (2,7) for $L=9$. 
We can expect that 
the configuration which appears with the highest probability 
is the ground state of the system, 
i.e., the vortex state with the lowest energy. 
(However, 
in some cases, i.e., for particular vortex configurations the highest probability 
state turns out to be not always the ground state configuration. 
One of these special cases will be addressed below.) 
This approach corresponds to the analysis done in the experiment
\cite{grigorieva}. 

The MD simulation was performed by using the Bardeen-Stephen
equation of motion
\begin{equation}
\eta \frac{d\rho_{i}}{dt}={\rm \bf F}_{i},
\label{bardeenstephen}
\end{equation}
where $i$ denotes the $i$th vortex, $\eta$ is the viscosity
coefficient
$\eta \sim \Phi_{0}H_{c2}/\rho_{n}c^{2}$,
with
$\rho_{n}$
being the normal-state resistivity; 
$\Phi_{0} = hc/2e$ is the magnetic flux quantum. 
The time integration was accomplished by using the Euler method.

\subsection{Vortex configurations for different $L$:
Formation of vortex shells}

\begin{figure}[btp]
\begin{center}
\hspace*{-0.5cm}
\includegraphics*[width=8.0cm]{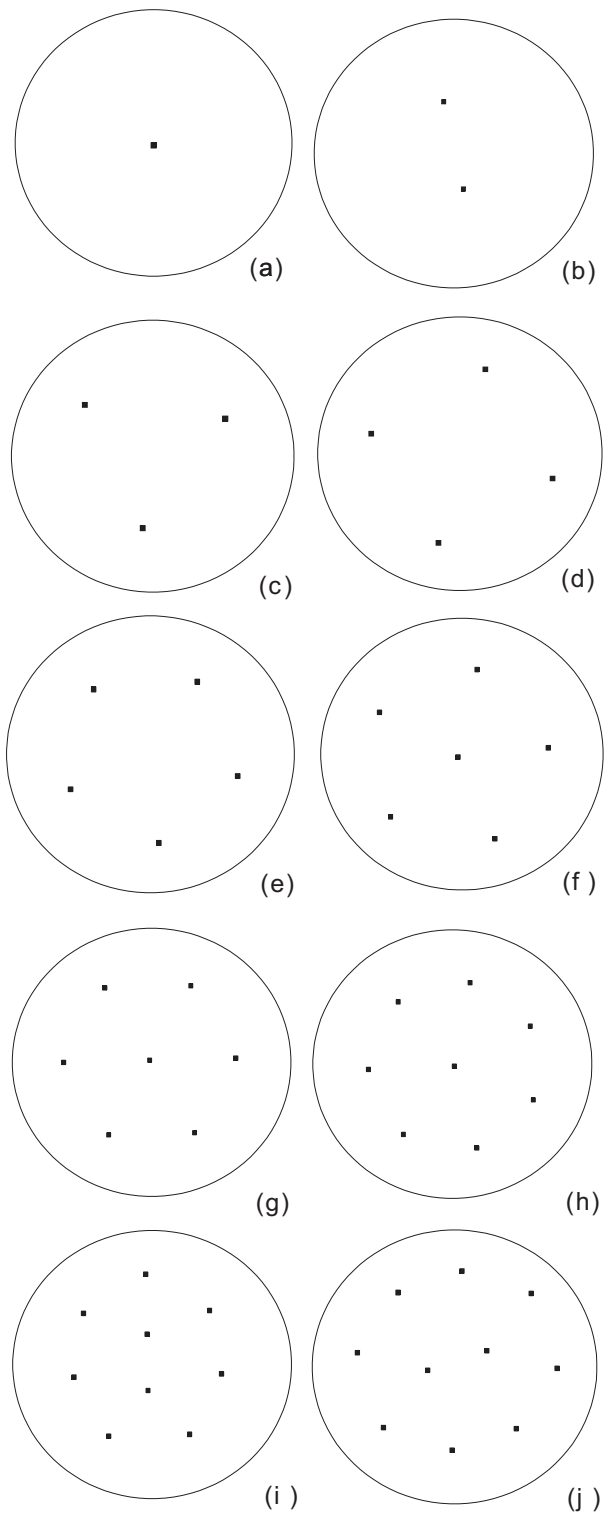}
\end{center}
\caption{
The evolution of vortex configurations for the states with vorticity
increasing from $L=1$ to 10, in mesoscopic superconducting disks 
with radius $R=50\xi$. Vortices form one-shell configurations for $L=1$ 
to 5 (a)-(e). The formation of second shell starts for $L=6$ (f). 
The inner shell contains 1 vortex at the center for $L=5$ to 8, while 
new vortices fill the outer shell (f-h). For $L=9$, second vortex
appears in the inner shell (i), and then second shell again starts
to grow for $L=10$ (j). } \vspace{-0.5cm}
\end{figure}

To study the formation of vortex shells in mesoscopic supeconducting 
disks, here we analyze the evolution of vortex configurations with increasing 
number of vortices, $L$, in a disk with radius $R=50\xi$. 
The results of our calculations for $L=1$ to 10 are presented in Fig.~1. 
When the vorticity $L$ of the sample increases,
the vortex configurations evolve with increasing applied magnetic
field as follows:
starting from a Meissner state without vortex,
then one appears in the center (Fig.~1(a)),
which we denote as (1),
then two symmetrically distributed vortices, (2) (Fig.~1(b)).
Further increasing the magnetic field
results in the formation of triangular, (3) (Fig.~1(c))
and square like (4)(Fig.~1(d)) vortex patterns in the sample,
and in a five-fold symmetric pattern, (5), shown in Fig.~1(e).
When the vorticity $L$ increases from 5 to 6, a vortex appears
in the center of the disk, thus starting to form a {\it second} shell 
of vortices in the disk (Fig.~1(f)). 
We denote the corresponding two-shell vortex state containing
1 vortex in the first shell and 5 vortices in the second shell
as (1,5).
Two-shell configurations with one vortex in the center and newly
generated vortices added to the outer shell remain for the
states (1,6) with $L=7$ (Fig.~1(g)), and (1,7) with $L=8$
(Fig.~1(d)).
The number of vortices in the inner shell begins to grow at
$L=9$ thus forming subsequently configurations (2,7)(Fig.~1(i)),
and (2,8) with $L=10$ (Fig.~1(j)).

Note that in earlier theoretical works on vortices in mesoscopic 
superconducting disks, a configuration (1,8) for $L=9$ was 
predicted \cite{BCPB04,lozovik} as a ground state in smaller disks, which however 
was not observed in the experiment \cite{grigorieva} as a stable state  
(the special case $L=9$ will be discussed in Sec.~III for the model of 
a {\it thick} disk, i.e., $d \gg \lambda$, relevant to the experiment 
\cite{grigorieva}, where the London screening in large disks, i.e., 
$R > \lambda$, is taken into account). 
Our calculations show that the multivortex states with two vortices 
in the center and the other vortices on the outer shell can exist 
till $L=14$. 
For $L>14$, the inner shell starts growing again till $L=16$,
which means that a newly nucleated vortex will be generated in the
inner shell, while the number of vortices in the outer shell stays
the same.
We found that those configurations are (3,11) for $L=14$, 
(4,11) for $L=15$, and (5,11) for $L=16$. 
At $L=17$, a vortex appears in the center, thus giving rise to the 
formation of a {\it third} shell with one vortex in the center. 
With increasing number of vortices in the disk, the next three vortices 
are added to the outermost shell, after which all three shells grow 
intermittently till $L=32$. 
The {\it fourth} shell appears at $L=33$ in the form of one vortex 
in the center. 
The borderline vortex configurations illustrating the formation of 
new shells are presented in Fig.~2. 
We summarize the vortex configurations found for vorticities 
$L = 1$ to 33 in Table~1.

\begin{figure}[btp]
\begin{center}
\hspace*{-0.5cm}
\includegraphics*[width=8.0cm]{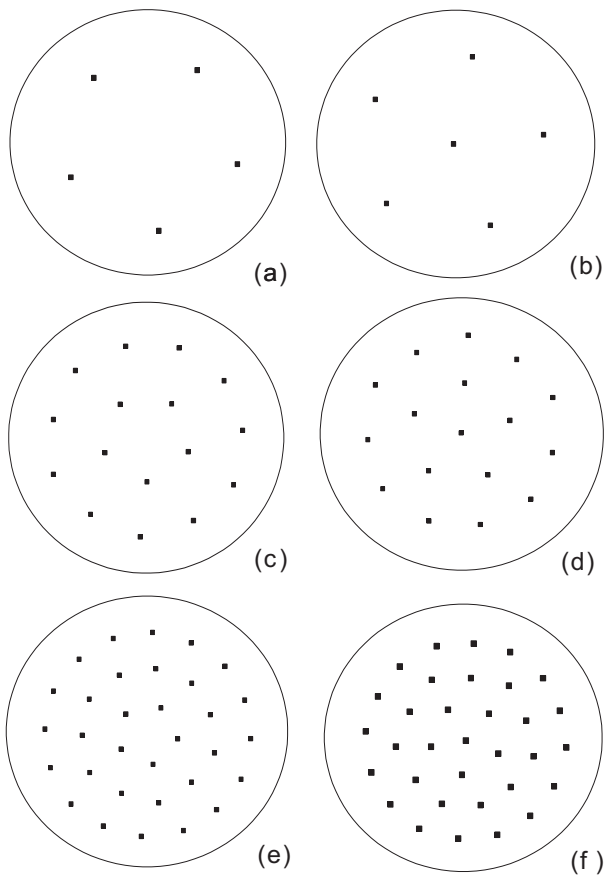}
\end{center}
\caption{
The borderline vortex configurations for: one shell, starting from
state (1) for $L=1$ to state (5) for $L=5$ (a), two shells, states
(1,5) for $L=6$ (b) to (5,11) for $L=16$ (c), three shells, states
(1,5,11) for $L=17$ (d) to (5,11,16) for $L=32$ (e), four shells,
starting from state (1,5,11,16) for $L=33$ (f), for disks with
radius $R=50\xi$. } \vspace{-0.5cm}
\end{figure}

It is appropriate to mention that in our numerical calculations using the 
vortex-vortex interaction force, Eq.~(\ref{fivv}), the obtained vortex patterns 
for some $L$ are different from those found in Ref.~\cite{lozovik} 
for particles with logarithmic interaction, confined by a parabolic potential, 
even in case when the interaction with images was taken into account. 
Although for many vorticities $L$ both approaches result in the same 
``robust'' configurations (i.e., less sensitive to details of interparticle 
interactions), there is essential difference in configurations for some 
other vorticities. 
The lowest vorticity for which our results deviate from those obtained 
in Ref.~\cite{lozovik} is $L=9$: this special case is described in detail 
below. 
Note, for instance, differences in three-shell configurations: 
(1,5,12) (our approach, see Table 1) [and (1,6,11) \cite{lozovik}] for $L=18$, 
(1,7,12) [(1,6,13)] for $L=20$, 
(1,8,13) [(1,7,14)] for $L=22$, 
(2,8,13) [(1,8,14)] for $L=23$, 
(3,10,15) [(4,9,15)] for $L=28$, 
(5,10,16) [(4,10,17)] for $L=31$, and 
(5,11,16) [(4,11,17)] for $L=32$. 
Moreover, the filling of next (fourth) shell starts, according to our calculations, 
for $L=33$ (i.e., (1,5,11,16)) while in Ref.~\cite{lozovik} this transition occurs 
for $L=34$.
As noticed in Ref.~\cite{lozovik}, the interaction with images leads to the 
stabilization of configurations with larger number of vortices on inner shells. 
While in \cite{lozovik} this tendency was revealed only in rather large clusters 
(i.e., the appearance of different configurations for $L=45$: i) (2,8,14,21), without 
interaction with images, and ii) (3,8,14,20), if the interaction with images 
is taken into account), we found that the stabilization of vortex shell structures 
with a larger number of vortices on inner shells occurs for vorticities $L \geq 20$, 
due to the interaction with images and with the shielding current induced by increasing 
magnetic field at the boundaries of the disk \cite{wecorbino}.

Note that, 
although here we employ the model of a thin mesoscopic disk, 
the results of our calculations for the filling of vortex shells 
perfectly match those discovered in the experiment \cite{grigorieva} 
(where the disks were rather thick) 
for vorticities $L=1$ to 40. 
The vortex configurations calculated in this section using many 
runs of simulations with random initial distributions, 
as will be shown below, 
are not always the ground-state configurations. 
The obtained results imply that the size of the disk 
influences the vortex configurations in superconducting disks. 
This will be clearly demonstrated in Sec.~III where we consider 
thick disks and we show that the London screening has a pronounced impact 
on the vortex configurations in disks of different radii.


\begin{table*}
%
\begin{tabular}{|c|c|c|c|c|c|c|c|c|c|c|c|}
\hline
   $L$     & 1 & 2 & 3 & 4 & 5 & 6 & 7 & 8 & 9 & 10 & 11 \\
\hline
   Config. & 1 & 2 & 3 & 4 & 5 & (1,5) & (1,6) & (1,7) & (2,7) & (2,8) & (3,8) \\
  \hline
   $L$     & 12 & 13 & 14 & 15 & 16 & 17 & 18 & 19 & 20 & 21 & 22 \\
\hline
   Config. & (3,9) & (4,9) & (4,10) & (4,11) & (5,11) & (1,5,11) & (1,5,12) & (1,6,12) & (1,7,12) & (1,7,13) & (1,8,13) \\
  \hline
   $L$     & 23 & 24 & 25 & 26 & 27 & 28 & 29 & 30 & 31 & 32 & 33 \\
\hline
   Config. & (2,8,13) & (2,8,14) & (3,8,14) & (3,9,14) & (3,9,15) & (3,10,15) & (4,10,15) & (4,10,16) & (5,10,16) & (5,11,16) & (1,5,11,16) \\
  \hline
\end{tabular}

\caption{\label{tab:table1} 
Formation of vortex shells in mesoscopic superconducting disks: 
vortex configurations for different vorticities $L$. 
}
\end{table*}


\subsection{The ground state and metastable states} 

As described in Sec.~II.A, in order to find the ground state 
of the system (or a state with energy very close to it), 
we performed many (usually one hundred) simulations 
for the same number of vortices. 
In most cases, always one configuration dominated over the other
possible configurations for a certain vorticity $L$, which was 
identified as the ``ground state''. 
However, for some vortex configurations competing states appeared with 
comparable probabilities. 

Let us now consider those special cases.
For instance, it follows from our calculations that
two configurations, (1,8) and (2,7), are possible for
the same vorticity $L=9$.
They are shown in Figs.~3(c),(d)
(another example of competing states with the same vorticity, e.g., $L=17$, are 
the three-shell configuration (1,5,11) and the two-shell configuration (5,12) 
which are shown in Figs.~3(a),(b), correspondingly). 
We found that the configuration (2,7) is the ground state
for $L=9$ in a disk with radius $R \leq 11\xi$ 
(see the $R-H$ phase diagram in Fig.~6) in a certain range of magnetic fields. 
In very large disks with $R \gtrsim 50\xi$ the vortex state (2,7), although being 
a metastable state, is the highest probability state. 
Note that configuration (2,7) was also found as a ground state for 
the system of charged particles \cite{BePB94}. 
At the same time, 
as it was shown in Ref.~\cite{BCPB04} using the GL theory, 
in small superconducting disks (e.g., for radius $R=6\xi$), 
this configuration occurred to be a metastable state, while state (1,8) 
was found as the ground energy state. 
This clearly shows that {\it the ground state configuration} 
for a certain $L$ {\it depends on the radius} of the disk.

\subsubsection{Statistical study of different vortex states}

In the experiments \cite{grigorieva}, vortex configurations were 
monitored in large arrays of similar mesoscopic disks (dots). 
This allowed them to study the statistics of the appearance of 
different vortex configurations in practically the same disk. 
The results show that, e.g., in a disk with radius $R=1.5\mu m$ and 
magnetic field $H_{0}=60 Oe$, configuration (2,8) for $L=10$ appeared 
more frequently. 
Other configurations for the same total vorticity $L=10$, e.g., 
configuration (3,7) appeared only in few cases. 
Interestingly, not only various configurations with the same total 
vorticity $L=10$ appeared, but also vortex states with $L=9$ (2,7) 
and --- less frequently --- 
two modifications of the state (1,8): with a ring-like outer 
shell 
as well as 
a square-lattice-like vortex 
pattern. 
This statistical study provides indirect information about the 
ground-state and metastable states.

\begin{figure}[btp]
\begin{center}
\hspace*{-0.5cm}
\includegraphics*[width=8.0cm]{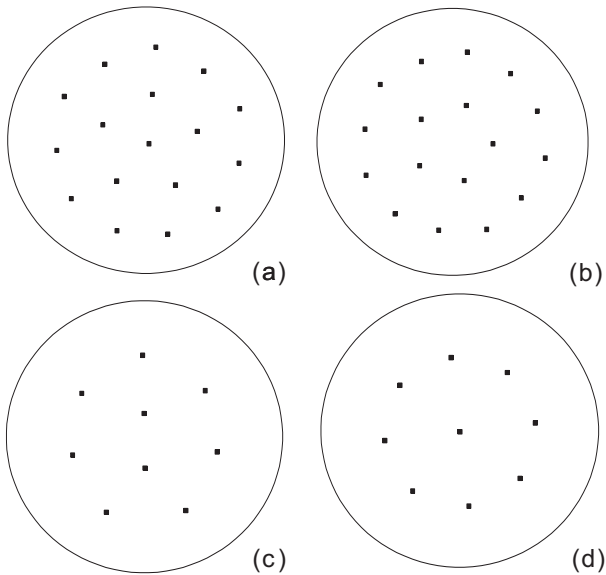}
\end{center}
\caption{
Possible vortex configurations for the total vorticity $L=17$ 
in a disk with radius $R=50\xi$: (1,5,11) (a) and (5,12) (b); 
and for the total vorticity $L=9$: (2,7) (c) and (1,8) (d).} 
\vspace{-0.5cm}
\end{figure}

We performed a similar investigation of the statistics of the appearance 
of different vortex states for ideal disks, i.e., in the absence of pinning. 
One hundred randomly distributed initial states were generated for our 
statistical study for each set of parameters. 
We studied the dependence of the appearance of different vortex 
configurations on the applied magnetic field. 
For instance, for a disk with radius $R=42\xi$ and the magnetic field 
varying from $H=0.011H_{c2}$ to 0.015$H_{c2}$, we counted how often the 
different configurations (e.g., (1,8) and (2,7)) for a total 
vorticity $L=9$ appeared.

\begin{figure}[btp]
\begin{center}
\vspace*{0.5cm}
\hspace*{-0.5cm}
\includegraphics*[width=8.0cm]{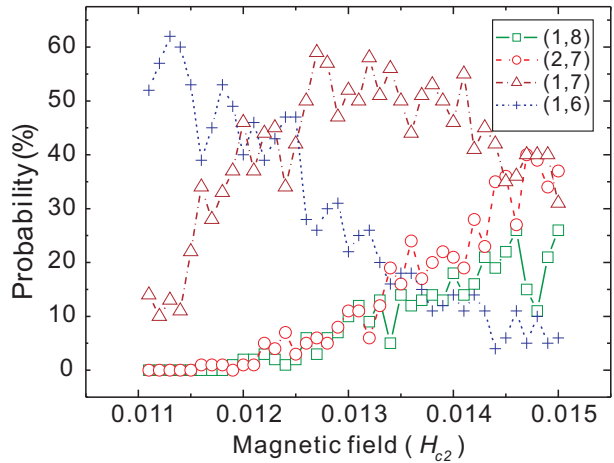}
\end{center}
\vspace{-0.5cm}
\caption{
The results of the statistical study of different vortex configurations, 
i.e., the probability to find a given vortex state as a function of the 
applied magnetic field, 
for a disk with radius $R=42\xi$ and varying magnetic field. 
}
\vspace{-0.5cm}
\end{figure}

The results of such calculations are shown in Fig.~4. 
At low magnetic field, $H=0.011H_{c2}$, the disk cannot accommodate 
9 vortices, so the number of configurations (1,8) and (2,7) is zero, 
and in most cases we obtain the configurations (1,6) or (1,7) for $L=7$ 
and $L=8$, respectively. 
As the magnetic field increases, the probability to find the configurations 
(1,8) and (2,7) increases, and at the same time 
the probability to find the configurations (1,6) and (1,7) decreases. 
Our statistical result shows clearly that 
in the range of magnetic field shown in Fig.~4 
the probability to find 
configuration 
(2,7) is always higher than to find configuration (1,8).

\begin{figure}[btp]
\begin{center}
\vspace*{0.5cm}
\hspace*{-0.5cm}
\includegraphics*[width=8.0cm]{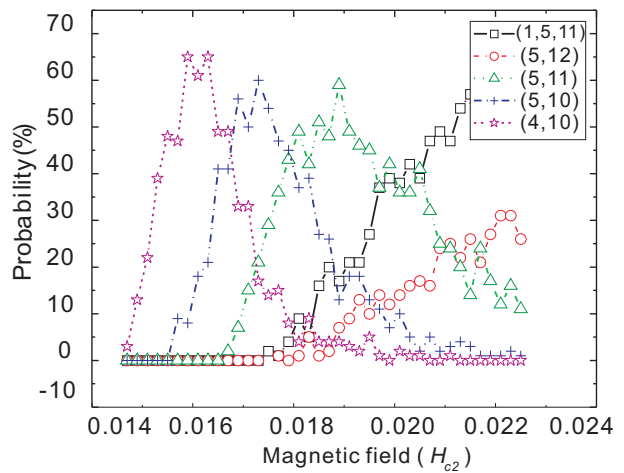}
\end{center}
\vspace{-0.5cm}
\caption{
The results of the statistical study of different vortex configurations, 
for a disk with radius $R=47\xi$ and varying magnetic field. 
} 
\vspace{-0.5cm}
\end{figure}

Similar analysis was performed for a disk with radius $R=47\xi$
where we looked for configurations with total vorticity 
$L=17$ in the range of the applied magnetic fields from
$H=0.014H_{c2}$ to 0.023$H_{c2}$.
As shown in Fig.~5,
configurations $(4,10)$ for $L=14$, $(5,10)$ for $L=15$,
$(5,11)$ for $L=16$, and $(1,5,11)$ and $(5,12)$ for $L=17$
dominate with increasing magnetic field.
Note that two configurations, (1,5,11) and (5,12), appeared 
in the same magnetic field range for $L=17$, and (1,5,11) is always the 
dominant configuration, 
i.e., 
the formation of the third shell starts for vorticity $L=17$
(cp. Ref.~\onlinecite{grigorieva}). 
The results of the statistical study of the configurations 
(2,7) and (1,8) for small disks ($R=9\xi$) are shown in Fig.~6.

\begin{figure}[btp]
\begin{center}
\vspace*{0.5cm}
\hspace*{-0.5cm}
\includegraphics*[width=8.0cm]{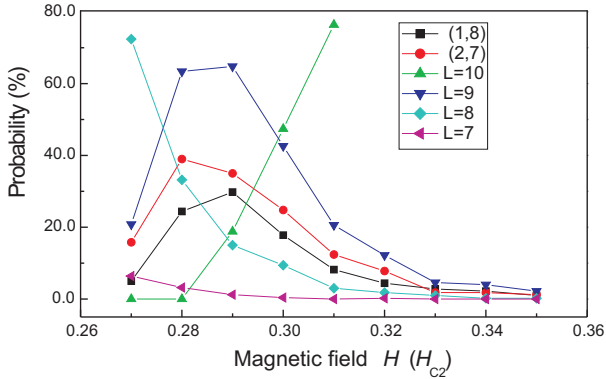}
\end{center}
\vspace{-0.5cm}
\caption{
The results of the statistical study of different vortex configurations, 
for a disk with radius $R=9\xi$ and varying magnetic field. 
The total probabilities of vortex states with different vorticities 
$L=7$, 8, 9, and 10, are plotted. 
For vorticity $L=9$, the probabilities of the two possible configurations 
(2,7) and (1,8) are also shown. 
} 
\vspace{-0.5cm}
\end{figure}

\subsubsection{The $R-H$ phase diagram}

To find the region of the existence and stability of vortex states 
with different vorticity $L$, 
we performed a direct calculation of their energies using Eq.~(\ref{fel}). 

As an example, we considered the configurations 
(1,8) and (2,7) in disks with radius changing 
in a very wide range from $R = 4\xi$ to $R \sim 100\xi$.  
We change the radius of the disk, and at the same time keep 
the flux passing through the disk $\Phi = S H_{0}$ the same, 
in order to keep the same vorticity $L$ in the disk. 
Here $\Phi$ is the flux passing through the specimen, $H_{0}$ is 
the applied magnetic field, $S=\pi R_{0}^{2}$ is the surface area of the specimen.

The phase diagram ``radius of the disk $R$ 
-- applied magnetic field $H$'' is shown in Fig.~7 
for $R=7\xi$ to $12\xi$. 
According to our calculations, 
for small radii $R<7\xi$, 
the energy of the configuration (2,7) appears to be always 
lower than that of configuration (1,8). 
The total energy for both configurations, (2,7) and (1,8), 
decreases with increasing 
magnetic field, 
and the energy of the state (2,7) is 
slightly larger than that of the state (1,8).

For disks with radius between $7\xi$ and $12\xi$,
the configuration (1,8) has a lower energy than configuration (2,7)
for low applied magnetic field.
For increasing magnetic field, 
the reverse is true. 
The rearrangement of the vortex configurations from the state (1,8) to 
the state (2,7) is related to the change 
in the steepness 
of the potential energy profile 
(i.e., the vortex-surface interaction, see Eq.~(\ref{epss})) 
for different points in the $R-H$ phase diagram. 
The inset of Fig.~7 shows the energy profiles corresponding to points 
C, D, and E in the phase diagram. 
Previously, it was shown for charged particles that the particle configiration 
is influenced by the steepness of the confinement potential \cite{kong}.

\begin{figure}[btp]
\begin{center}
\vspace*{0.5cm}
\hspace*{-0.5cm}
\includegraphics*[width=8.0cm]{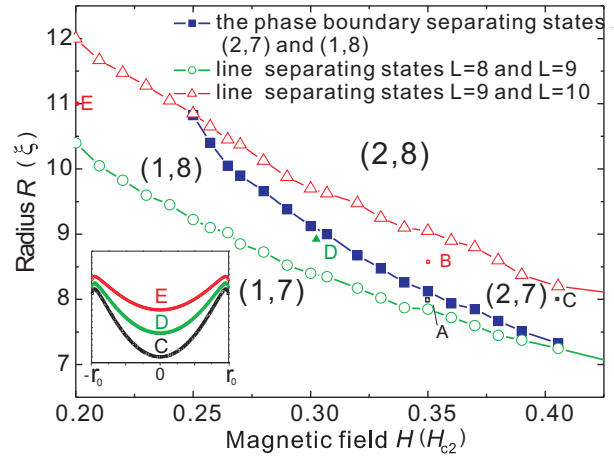}
\end{center}
\vspace{-0.5cm}
\caption{
The $R-H$ phase diagram for states with total vorticity $L=9$, 
for small disks with radii changing in the range: $R=7\xi$ to $R=12\xi$. 
Points A and B correspond to the configurations (a) and (b) shown in Fig.~10, 
respectively. 
The inset shows the confinement potentials (see Eq.~(\ref{epss})) corresponding to the points 
C, D, and E in the phase diagram. 
}
\vspace{-0.5cm}
\end{figure}

As it follows from the phase diagram (Fig.~7), 
we can expect that for larger radii 
(in the mesoscopic regime, i.e., when $R < \lambda_{eff}$; 
we show in Sec.~III that for thick disks with $R > \lambda_{eff}$
the configuration (2,7) restores as the ground-state configuration). 
the configuration (1,8) has a lower energy than 
the state (2,7), in the low magnetic field range. 
However, the difference in energy between the states (2,7) and (1,8) 
decreases for increasing $R$, 
although the energy of the state (1,8) remains lower than that 
of the state (2,7) for even larger disks 
(e.g., with radius $R \sim 40-90\xi$ but still ``mesoscopic'' due 
to the condition $R < \lambda_{eff}$ valid for very thin disks). 
This seems to contradict our previous result that the state (2,7) 
is the {\it highest probability} state in large disks (which is also 
in agreement with the experiment \cite{grigorieva}). 
Thus the question is, 
whether the highest probability configuration (e.g., state (2,7)) is always 
the ground state? If not, what is the reason for that?

\begin{figure}[btp]
\begin{center}
\vspace*{0.5cm} \hspace*{-0.5cm}
\includegraphics*[width=8.5cm]{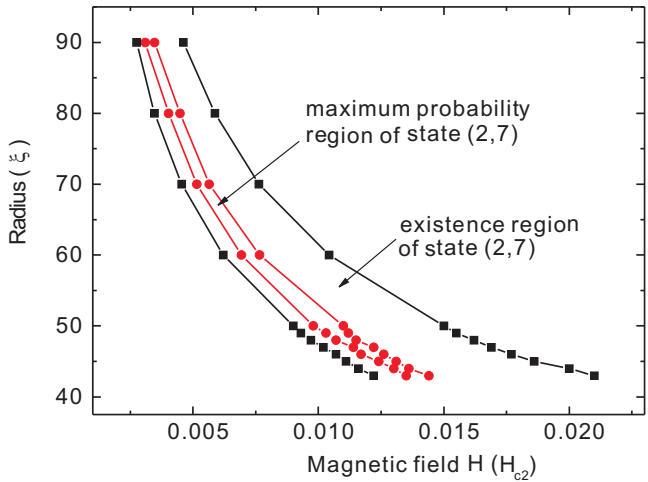}
\end{center}
\vspace{-0.5cm} 
\caption{ 
The $R$-$H$ diagram for the vortex state (2,7) 
calculated using the statistical approach, 
for large disks with radius $R=40\xi$ to $90\xi$.
The area between the red curves and dots shows the region where 
the state (2,7) has the maximum probability. 
} 
\vspace{-0.5cm}
\end{figure}

To answer this question, we calculated, 
using the statistical approach described above, 
the $R-H$ diagram for the state (2,7) (Fig.~8), 
i.e., the region of the existence of the state (2,7) 
and the region where the state (2,7) has the highest probability, 
for radii $R=40\xi$ to $90\xi$. 
The region of the existence of the state (2,7) as the highest probability state 
is very narrow although it is well-defined even for very large radii 
(e.g., $R=90\xi$, see Fig.~8). 
However, for radii $R \lesssim 40\xi$ this region becomes narrower and 
unstable, i.e., it can even disappear, giving rise to higher probability 
of appearance of the state (1,8). 

This calculation clearly shows that the highest probability state (2,7) 
is {\it not} the ground state in large disks. 
The reason of such behavior is that the energy minimum in configurational space 
corresponding to the state (2,7) is {\it very wide} while the competing state (1,8) 
possesses, although slightly deeper, a {\it much narrower} minimum. 
Thus, statistically the system ends up more often into the wide minimum corresponding 
to the state (2,7). 
This is confirmed by a calculation of the potential energy profile which is shown 
in Fig.~9 as a function of the displacement 
(i.e., for different vortex configurations, between the 
initial nonequilibrium configuration (2,7) -- through the equilibrium state (2,7) -- 
and the final state (1,8)) 
of one of the central vortices of the configuration (2,7) 
during the continuous transition to the configuration (1,8). 
The position of the other vortices is determined by minimizing the energy. 
The corresponding changes of the vortex configuration are shown in the inset 
of Fig.~9. 
We started from an out-of-equilibrium (2,7) configuration, 
passed through the equilibrium (2,7) configuration (wide minimum), 
then passed over the energy maximum, and, 
finally, ended up at the equilibrium (1,8) configuration (a tail of the 
transition is also shown for out-of-equilibrium (1,8) configuration) 
which has slightly lower energy.

\begin{figure}[btp]
\begin{center}
\hspace*{-0.5cm}
\includegraphics*[width=8.5cm]{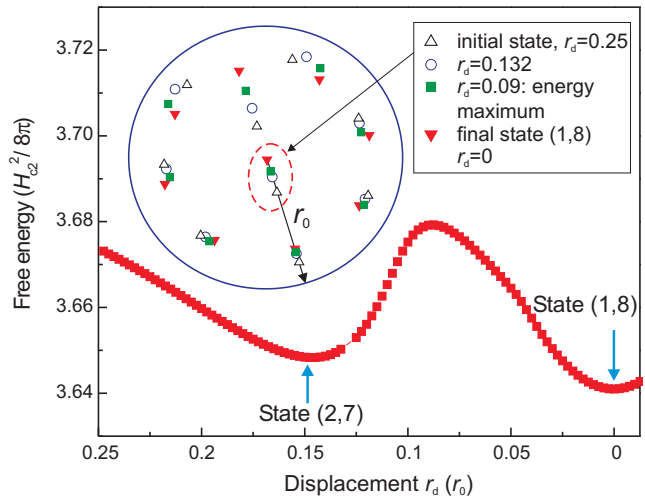}
\end{center}
\vspace{-0.5cm} 
\caption{ 
The energy of the system of $L=9$ vortices 
as a function of the displacement $r_{d}$ 
(measured in units of the disk radius $r_{0}$ and counted from the center of the disk) 
of one of the central vortices of the initial configuration (2,7) 
during the continuous transition to the configuration (1,8). 
The inset shows the corresponding evolution of the vortex configuration 
(different symbols show vortex configurations for different dispalcements $r_{d}$ 
of the vortex marked by red dashed line) 
during the 
transition from the state (2,7) to the state (1,8). 
} 
\vspace{-0.5cm}
\end{figure}

\section{Macroscopic regime: Thick disks} 

In Sec.~II we analyzed in detail the formation of vortex shells in 
mesoscopic disks. 
Although we considered rather large disks with radii up to 
$R \sim 100\xi$, the results obtained in the previous section 
refer essentially to {\it thin} disks: only if the disk's thickness $d$ 
is small enough, the $R < \lambda_{eff}=\lambda^{2}/d$ condition 
is satisfied, i.e., for disks with $R \sim 1\mu$m, $d$ has to be of 
the order of a few nanometers. 
Nb disks used in the experiment \cite{grigorieva} had radius $R = 1-2\mu$m 
and thickness $d = 150$nm. 
For such {\it thick} disks with $R > \lambda$ (e.g., $\lambda (0) = 90$nm 
for Nb disks in the experiment \cite{grigorieva}), the effects due to London 
screening become important. 
In this section, we consider the limit of thick disks with $d \gg \lambda$ 
and we study how the London screening in the vortex-vortex and 
vortex-boundary interactions influences the vortex patterns in the disk. 


Here we model a cylinder with radius $R$ infinitely long in the $z$-direction 
by a two-dimensional (2D) (in the $xy$-plane) disk, 
assuming the vortex lines are parallel to the cylinder axis. 
This approach was used for studying, e.g., vortex dynamics in periodic 
\cite{md01,md0157,md02} and quasiperiodic arrays of pinning sites (APS) 
\cite{penrose}. 
As distinct from infinite APSs where periodic boundary 
conditions are imposed at the boundaries of a simulation cell, 
here we impose boundary conditions at the edge of the (finite-size) disk, 
namely, the potential barrier for vortex entry/exit. 
To study the configurations of vortices interacting with each other 
and with the potential barrier, 
we perform simulated annealing simulations by numerically integrating 
the overdamped equations of motion 
(unlike in Refs.~\cite{md01,md0157,md02,penrose}, there is no external 
driving force in our system, and we study the relaxation of initially 
randomly distributed vortices to the ground-state vortex configuration): 
\begin{equation}
\eta {\rm \bf v}_{i} \ = \ {\rm \bf f}_{i} \ = \ {\rm \bf f}_{i}^{vv} + 
{\rm \bf f}_{i}^{vb} + {\rm \bf f}_{i}^{T}. 
\label{eqmd} 
\end{equation} 
Here, 
${\rm \bf f}_{i}$ 
is the total force per unit length acting on vortex 
$i$, 
${\rm \bf f}_{i}^{vv}$ 
and 
${\rm \bf f}_{i}^{vb}$ 
are the forces due to vortex-vortex and vortex-barrier interactions, 
respectively, 
and 
${\rm \bf f}_{i}^{T}$ 
is the thermal stochastic force; 
$\eta$ is the viscosity, which is set here to unity. 
The force due to the interaction of the $i$-th vortex with other vortices is 
\begin{equation}
{\rm \bf f}_{i}^{vv} \ = \ \sum\limits_{j}^{N_{v}} \ f_{0} \ K_{1} \!
\left( \frac{ \mid {\rm \bf r}_{i} - {\rm \bf r}_{j} \mid }{\lambda} \right)
\hat{\rm \bf r}_{ij} \; , 
\label{fvv}
\end{equation}
where 
$N_{v}$ 
is the number of vortices, 
$K_{1}$
is the modified Bessel function, 
$\hat{\rm \bf r}_{ij} = ( {\rm \bf r}_{i} - {\rm \bf r}_{j} )
/ \mid {\rm \bf r}_{i} - {\rm \bf r}_{j} \mid,$
and 
$$ 
f_{0} = \frac{ \Phi_{0}^{2} }{ 8 \pi^{2} \lambda^{3} } \; . 
$$ 
It is convenient, following the notation used in 
Refs.~\cite{md01,md0157,md02,penrose}, 
to express now all the lengths in units of 
$\lambda$ 
and all the fields in units of 
$\Phi_{0}/\lambda^{2}$. 
The Bessel function 
$K_{1}(r)$ 
decays exponentially for 
$r$
larger than
$\lambda$, 
therefore it is safe to cut off the (negligible) force for distances larger 
than $5\lambda$. 
In our calculations, the logarithmic divergence of the vortex-vortex interaction 
forces for 
$r \to 0$
is eliminated by using a cutoff for distances less than
$0.1\lambda$. 

Vortex interaction with the edge is modelled by implying the usual Bean-Levingston 
barrier \cite{bean-levingston,tinkham,book}. 
We assume that the repulsive force exerted by the surface current on the vortex 
at a distance $r$ from the disk edge decays as 
\begin{equation}
{\rm \bf f}_{i}^{vbc} = \frac{ \Phi_{0} H_{0} }{ 4 \pi \lambda } {\rm exp }
\left( -\frac{r}{\lambda} \right), 
\label{vbc}
\end{equation}
as it does in the case of a semi-infinite superconductor 
\cite{bean-levingston,tinkham,book} 
(which is justified for disks with $R \gg \lambda$), 
and the attractive force due to the vortex interaction 
with its image is expressed by
\begin{equation}
{\rm \bf f}_{i}^{vbi} = - {\rm \bf f}_{i}^{vv}, 
\label{vbi}
\end{equation}
and 
\begin{equation}
{\rm \bf f}_{i}^{vb} = {\rm \bf f}_{i}^{vbc} + {\rm \bf f}_{i}^{vbi}. 
\label{vb}
\end{equation}
Here we assume that for large enough disks, the distance from the edge to the image 
is equal to the distance to the vortex. 

The temperature contribution to Eq.~(\ref{eqmd}) is represented by a stochastic 
term obeying the following conditions: 
\begin{equation}
\langle f_{i}^{T}(t) \rangle = 0, 
\end{equation}
and
\begin{equation}
\langle f_{i}^{T}(t)f_{j}^{T}(t^{\prime}) \rangle = 2 \,  \eta \,  k_{B} \,  T \,  \delta_{ij} \,  
\delta(t-t^{\prime}). 
\end{equation}

The ground state of a system of vortices is obtained as follows. 
First, we set a high value for the temperature, to let vortices move randomly. 
Then, the temperature is gradually decreased down to $T = 0$. 
When cooling down, vortices interacting with each other and 
with the edges adjust themselves to minimize the energy, 
simulating the field-cooled experiments (see, e.g., \cite{tonomura-vvm,togawa}).


Our calculations show that most of the vortex configurations found 
in Sec.~II and in the previous theoretical works on mesoscopic 
disks \cite{CBPB04} remain unchanged also in large disks where the 
interactions are screened at the London penetration depth $\lambda$. 
These are stable shell patterns (e.g., (1,6) for $L=7$ and 
(1,7) for $L=8$, etc.) which were found to be the ground-state 
configurations of vortices in superconductors \cite{CBPB04}, 
in liquid He \cite{campbell}, and in a system of charged particles 
confined by a parabolic potential \cite{BePB94}. 
These stable configurations are mainly determined by the circular 
shape of the disk, and they are to a much less extend sensitive to 
the specific interaction potentials between the particles and 
the boundaries. 
On the other hand, 
the ``borderline'' configurations (i.e., those for which one or 
more shells start to be filled), e.g., the states (1,8) vs. (2,7) 
for $L=9$ or the states (2,9) vs. (3,8) for $L=11$, are much more 
sensitive to the interactions in the disk. 
For example, for $L=11$ the theory predicts the configuration 
(3,8) to be the ground state for vortices in He \cite{campbell} 
and for charged particles \cite{BePB94}, and it is was also 
observed in the experiment \cite{grigorieva} in {\it large} disks, 
while the theory predicted \cite{CBPB04} the configuration (2,9) 
in {\it small} mesoscopic disks. 
For $L=9$ the theory predicts that the configuration (1,8) 
is the ground state for vortices in He \cite{campbell} and in small 
mesoscopic superconducting disks \cite{CBPB04}, while for charged 
particles the configuration (2,7) was predicted \cite{BePB94}. 
This vortex configuration, (2,7), was also observed in the 
experiment \cite{grigorieva} with Nb disks. 
In Sec.~II we showed that in mesoscopic disks the state (2,7) 
had the highest probability to appear (due to the wide potential 
energy minimum related to this state), although it was not the 
lowest energy configuration, but instead the (1,8) configuration 
was the ground state.

\begin{figure}[btp]
\begin{center}
\hspace*{-0.5cm}
\includegraphics*[width=7.5cm]{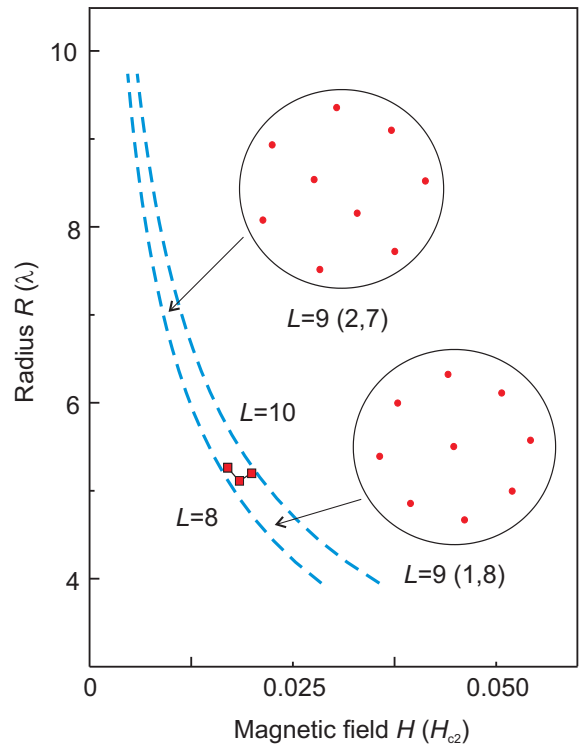}
\end{center}
\vspace{-0.5cm} 
\caption{ 
The $R-H$ phase diagram for vortex states with total 
vorticity $L=9$ in case of a thick disk. 
The 
area between the 
dashed blue curves shows the region where the 
states with $L=9$ are the ground state. 
The boundary separating the states (1,8) (small radii) 
and (2,7) (large radii) are shown by solid red squares. 
The insets show the corresponding vortex patterns, (1,8) and (2,7). 
} 
\vspace{-0.5cm}
\end{figure}

In case of thick disks as considered here, the calculations show 
the {\it crossover} behavior of the vortex patterns from the 
state (1,8) to (2,7) with increasing radius of the disk. 
The phase diagram in Fig.~10 illustrates this behavior. 
Note that the potential barrier at the disk edge becomes extremely 
low for low values of the applied magnetic field $H_{0}$ when we have 
a large radius of the disk. 
This means that it is very difficult to stabilize a vortex state 
with only {\it few} vortices in such a large disk (in experiment, 
and also in the numerical calculations using the Ginzburg-Landau 
equations) because for even very low barrier at the boundary 
{\it many} vortices can enter the sample without any appreciable 
change of the flux inside the disk. 
The lines separating the states with different vorticities (shown 
by dashed lines in Fig.~10) are calculated here assuming the flux 
inside the disk is on average equal to the applied magnetic field 
$H_{0}$ multiplied by the area of the disk. 
The calculated line separating the states (1,8) and (2,7) is shown 
by solid squares. 
The phase diagram shows that in relatively small disks with radius 
$R \lesssim 5 \lambda$ (that is $R \lesssim 30 \xi$ in case of Nb 
disks \cite{grigorieva}), the configuration (1,8) is the ground state 
while for larger disks we find the state (2,7), in agreement with 
the experiment \cite{grigorieva}.

\begin{figure}[btp]
\begin{center}
\hspace*{-0.5cm}
\includegraphics*[width=7.0cm]{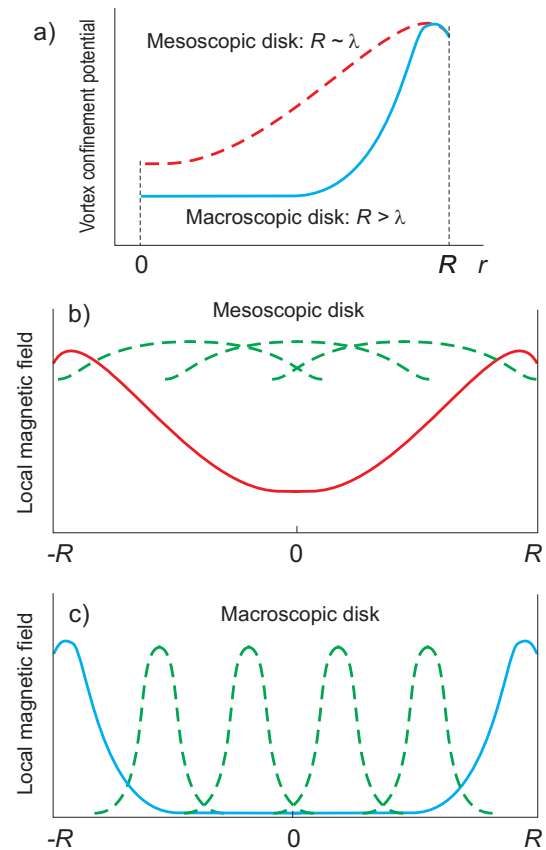}
\end{center}
\vspace{-0.5cm} 
\caption{ 
The vortex confinement energy profiles for: 
mesoscopic disk ($R \sim \lambda$) 
and for 
macroscopic disk ($R > \lambda$) (a). 
Schematic plots illustrating the magnetic field profiles 
(green dashed curves) 
of the interacting vortices in: 
mesoscopic disk (solid red curve shows the vortex confinement potential) 
(b), 
and in 
macroscopic disk (solid blue curve shows the vortex confinement potential) 
(c). 
} 
\vspace{-0.5cm}
\end{figure}

This crossover behaviour could be understood in the following way. 
In Fig.~11(a) we plot the vortex confinement potential profiles for 
a mesoscopic disk ($R \sim \lambda$) and for a macroscopic disk 
($R > \lambda$). 
In a mesoscopic disk, {\it all} the vortices interact with the 
screening current which extends inside the disk. 
In a macroscopic disk, {\it only the outer-shell} vortices feel the 
screening current. 
More importantly, the intervortex interaction changes in a disk 
with the London screening: 
in a mesoscopic disk, 
{\it each vortex interacts with all other vortices} 
since the currents created by the vortices strongly overlap 
(see Fig.~11(b)), 
and the minimum potential energy is reached when the 
{\it sum of all the intervortex distances is maximum}, 
i.e., for the configuration (1,8). 
In a macroscopic disk, 
the intervortex interaction is very weak, and 
{\it each vortex interacts only with its closest neighbor} 
through the tails of the currents associated with each vortex 
(see Fig.~11(c)), 
and the minimum potential energy is reached when the 
{\it sum of closest-neighbor intervortex distances is maximum}, 
i.e., for the configuration (2,7). 
The vortex pattern (2,7) in a large disk (see inset in Fig.~10) 
resembles a distorted Abrikosov vortex lattice in an infinite 
superconductor which is stabilized by intervortex interactions 
in the absence of boundaries (note that the outer-shell vortices 
are relatively closer to the boudary and the two vortices in the 
inner shell are slightly out of the center minimizing the 
interaction energy with the 2 and 3 neighbors). 

The calculated crossover behavior found here is consistent with  
the $R-H$ phase diagram obtained in Sec.~II for mesoscopic disks 
(Fig.~7) that predicted the configuration of (1,8) as the ground 
state for radii $R \gtrsim 10\xi$. 
Thus, according to the phase diagrams for mesoscopic disks (Fig.~7) 
and for macroscopic disks (Fig.~10), there are {\it two} crossovers 
between the states (1,8) and (2,7): 
the configuration (1,8) is the ground state in disks with radius 
$10\xi \lesssim R \lesssim 30\xi$, while 
the configuration (2,7) occurs to be the ground state in large disks 
with $R \gtrsim 5 \lambda$ (i.e., $R \gtrsim 30 \xi$ in Nb disks 
\cite{grigorieva}) {\it and} in very small disks with $R \lesssim 10 \xi$. 
The mechanism of the second crossover for very small disks is very 
different from that for large disks, and the transition (1,8)$\to$(2,7) 
happens in very small disks due to a strong overlap of the vortex cores 
in the outer eight-vortex shell: the vortices cannot accomodate on the 
outer shell and one of them is pushed towards the interior of the disk 
(note that for even smaller disks the configuration (2,7) collapses to 
a giant-vortex state). 
This behavior will be demonstrated in Sec.~IV using the Ginzburg-Landau 
theory.

\section{Comparison with the GL theory}

In order to go beyond the London approximation we used also the 
Ginzburg-Landau (GL) equations to calculate the free energy and 
find the ground state.
Within the GL approach vortices are no longer point-like ``particles''
but extended objects.
The expression for the dimensionless Gibbs free energy is
(see, e.g., Ref.~\cite{SPB98}): 
\begin{equation}
G = V^{-1}\int_{V}[2({\rm \bf A}-{\rm \bf A}_{0})
\cdot {\rm \bf j}_{2D}- \mid \Psi \mid^{4}]d {\rm \bf r},
\label{}
\end{equation}
with $\psi({\rm \bf r})$ the order parameter, 
${\rm \bf A} ({\rm \bf A}_{0})$ the vector potential 
of the total (applied) magnetic field and ${\rm \bf j}_{2D}$ 
the superconducting current. 
By comparing the dimensionless Gibbs free energies of the 
different vortex configurations, we find the ground state. 
Similarly we could find the two stable configurations (2,7) and 
(1,8) in a disk with vorticity $L=9$ as we found within the 
MD simulations in Secs~II and III.

\begin{figure}[btp]
\begin{center}
\vspace*{0.5cm}
\hspace*{-0.5cm}
\includegraphics*[width=8.5cm]{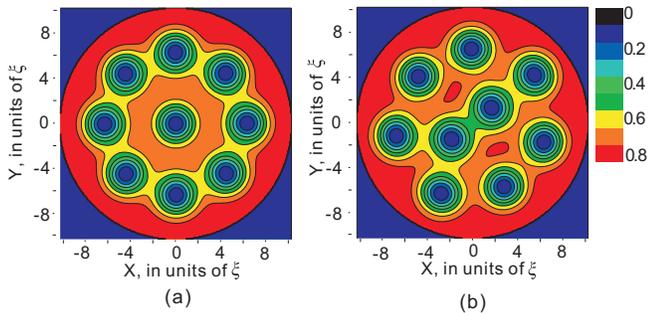}
\end{center}
\vspace{-0.5cm}
\caption{
Contour plots of the Cooper-pair density of state (2,7) for $R=8.3\xi$ (a), 
and state (1,8) for $R=8.2\xi$ (b), in an applied magnetic field $H_{0}=0.35H_{c2}$. 
Blue and red regions correspond to low and high Cooper-pair density, respectively. 
}
\vspace{-0.5cm}
\end{figure}

The results of our calculations of the order parameter distribution 
using the GL equations (for simplicity, this was done for 
zero disk thickness, $d \to 0$, 
i.e., in the limit of extreme type II superconductor, for a given 
applied magnetic field, i.e., only first GL equation was solved)
for the total vorticity $L=9$ are shown in Fig.~12. The states (2,7) and
(1,8) are shown in the phase diagram (see Fig.~7) by symbols A and B, 
respectively. 
Samples with 
different radius were examined for a fixed external magnetic field
$H=0.35H_{c2}$. For a disk with radius $R=8.2\xi$, our calculation
gives (1,8) as the ground vortex state. When the radius of the disk 
is increased, the energy of (1,8) is the lowest one till $R=8.25\xi$, 
after which configuration (2,7) becomes the ground state. 
This compares with our results of Sec.~II using the London theory where 
we found that the transition (1,8) $\rightarrow$ (2,7) occurred for $R=8.13\xi$ 
when $H=0.35H_{c2}$. 
Thus, the results of a calculation of the vortex 
configurations within the GL, with an appropriate choice of the radius 
and external parameters, confirms the crossover behavior found in Sec.~II.

\section{Conclusions}


In this work, we studied the vortex configurations in mesoscopic 
superconducting thin disks 
and in thick disks taking into account the London screening,  
using the 
Molecular-Dynamics simulations of the Langevin-type equations of 
motion 
and confirmed these results, in case of small disks, using the more extended 
Ginzburg-Landau functional theory. 

This study was motivated by recent experiments 
by Grigorieva {\it et al.} \cite{grigorieva} 
who observed vortex {\it shell} structures in mesoscopic Nb 
disks with $R \sim 1-2.5\mu$m by means of the Bitter decoration technique. 
It was shown in those experiments, that in disks with vorticity 
ranging from $L = 1$ to 40, vortices fill the disk according 
to specific rules, forming well-defined shell structures, 
as earlier predicted in Ref.~\cite{BePB94}. 
They analyzed the formation of these shells which resulted 
in a ``periodic table'' of formation of shells. 
It was shown that most of the experimentally observed
configurations for small $L$ agreed with those theoretically
predicted earlier \cite{BCPB04,BePB94}. 
At the same time, some of the configurations which were observed
in these experiments were not found earlier in vortex systems
(although they were shown to appear in  systems of charged particles
and in superfluids).

In this work, 
we found the rules according to which the shells are filled with
vortices for increasing applied magnetic field.
In particular, it was shown in our calculations, that for the
vortex configurations with the number of vortices up to $L = 5$,
the vortices form a single shell. 
The formation of a second shell starts from $L = 6$.
Similarly, the formation of a third shell starts at $L = 17$, and
of a fourth shell at $L = 33$.
These theoretical findings are in agreement with the results of
the experimental observations of Ref.~\cite{grigorieva}.
Moreover, we found those states which were observed in the experiments 
but not found in previous calculations. 
Thus, we filled the missing states in the ``periodic table'' 
of vortex shells in mesoscopic disks. 
We studied in detail the region of parameter space where those states 
exist, and compared the obtained results to previous theoretical works 
where small mesoscopic disks with $R \sim 5-10 \xi$ were considered. 

It was shown that some of the vortex configurations 
(i.e., those which are at the borderline between configurations 
characterized by different stable shell structures) 
are very sensitive to the size of the disk. 
For instance, 
we found that depending on the radius of the disk, 
there are two crossovers between the states (1,8) and (2,9) for $L=9$: 
at $R \sim 10\xi$ and $R \sim 30\xi$. 
The 
(1,8) $\rightarrow$ (2,7) 
transition occurs for disks with $R \sim 5\lambda$ 
(that corresponds to $R \sim 30\xi$ in case of Nb disks in the 
experiment \cite{grigorieva}) 
due to the effect of the London screening in large disks, 
while in small disks with $R \sim 10\xi$ 
this transition happens due to the compression of the outer 
eight-vortex shell. 

Thus we performed a systematic study of the size-dependence 
of vortex configurations in mesoscopic superconducting disks. 
Our results agree with the experimental observations of vortex 
shells in Nb disks \cite{grigorieva} 
and explain the revealed discrepancies with the 
earlier calculations of vortex shells.

\section{Acknowledgments} 

We thank D.Yu.~Vodolazov, M.V.~Milo\v{s}evi\'{c}, and B.J.~Baelus for useful 
discussions.
This work was supported by the Flemish Science Foundation (FWO-Vl) and the 
Interuniversity Attraction Poles (IAP) Programme –- Belgian State –- Belgian Science 
Policy. V.R.M. acknowledges partial support through POD.


\begin{references}

\bibitem[$^{\ast}$]{A1}
Electronic address: francois.peeters@ua.ac.be 

\bibitem{GPN97}
A.K. Geim, I.V. Grigorieva, S.V. Dubonos, J.G.S. Lok, J.C. Maan, A.E. Filippov, 
and F.M. Peeters, Nature (London) {\bf 390}, 256 (1997). 

\bibitem{DSPGL97}
P.S. Deo, V.A. Schweigert, F.M. Peeters, and A.K. Geim,
Phys. Rev. Lett. {\bf 79}, 4653 (1997).

\bibitem{lozovik}
Yu.E.~Lozovik and E.A.~Rakoch, Phys. Rev. B {\bf 57}, 1214 (1998).

\bibitem{SPB98}
V.A.~Schweigert and F.M.~Peeters, Phys. Rev. B {\bf 57}, 13817 (1998).

\bibitem{SPL99}
V.A. Schweigert and F.M. Peeters, Phys. Rev. Lett. {\bf 83}, 2409 (1999).

\bibitem{BPS01}
B.J.~Baelus, F.M.~Peeters, and V.A.~Schweigert, Phys. Rev. B {\bf 63},
144517 (2001).

\bibitem{tinkham}
M.~Tinkham, {\it Introduction to superconductivity},
(McGraw-Hill, New York, 1996), 2nd ed.

\bibitem{degennes}
P.G.~de~Gennes, {\it Superconducting of Metals and Alloys},
(Benjamin, New York, 1966).

\bibitem{abrikosov}
A.A.~Abrikosov, {\it Fundamentals of the Theory of Metals},
(North-Holland, Amsterdam, 1986).

\bibitem{buzdin}
A.I.~Buzdin and J.P.~Brison, Phys. Lett. A {\bf 196}, 267 (1994).

\bibitem{palacios}
J.J.~Palacios, Phys. Rev. B {\bf 58}, R5948 (1998).

\bibitem{geim}
A.K.~Geim, S.V.~Dubonos, J.J.~Palacios, I.V.~Grigorieva, M.~Henini,
and J.J.~Schermer, Phys. Rev. Lett. {\bf 85}, 1528 (2000).

\bibitem{MYPB02}
M.V.~Milo\v{s}evi\'{c}, S.V.~Yampolskii, and F.M.~Peeters,
Phys. Rev. B {\bf 66}, 024515 (2002).

\bibitem{MPB03}
M.V.~Milo\v{s}evi\'{c} and F.M.~Peeters,
Phys. Rev. B {\bf 68}, 024509 (2003).

\bibitem{BCPB04}
B.J.~Baelus,
L.R.E.~Cabral, and F.M.~Peeters, 
Phys. Rev. B {\bf 69}, 064506 (2004).

\bibitem{grigorieva}
I.V.~Grigorieva,
W.~Escoffier, J.~Richardson, L.Y.~Vinnikov, S.~Dubonos, and V.~Oboznov,
Phys. Rev. Lett. {\bf 96}, 077005 (2006).

\bibitem{campbell}
L.J.~Campbell and R.M.~Ziff, Phys. Rev. B {\bf 20}, 1886 (1979).

\bibitem{hess}
G.B.~Hess, Phys. Rev. {\bf 161}, 189 (1967). 

\bibitem{stauffer}
D.~Stauffer and A.L.~Fetter, Phys. Rev. {\bf 168}, 156 (1968). 

\bibitem{totsuji}
H.~Totsuji and J.L.~Barrat, Phys. Rev. Lett. {\bf 60}, 2484 (1988). 

\bibitem{tsuruta}
K.~Tsuruta and S.~Ichimaru, Phys. Rev. A {\bf 48}, 1339 (1993). 

\bibitem{BePB94}
V.M.~Bedanov and F.M.~Peeters, Phys. Rev. B {\bf 49}, 2667 (1994). 

\bibitem{plasma} 
Y.-J.~Lai and L.~I, Phys. Rev. E {\bf 60}, 4743 (1999). 

\bibitem{colloids} 
I.V.~Schweigert, V.A.~Schweigert, and F.M.~Peeters, 
Phys. Rev. Lett. {\bf 84}, 4381 (2000); 
K.~Mangold, J.~Birk, P.~Leiderer, and C.~Bechinger, 
Phys. Chem. Chem. Phys. {\bf 6}, 1623 (2004). 

\bibitem{kong} 
M.~Kong, B.~Partoens, and F.M.~Peeters, Phys. Rev. E {\bf 65}, 046602 (2002). 

\bibitem{CBPB04}
L.R.E.~Cabral,
B.J.~Baelus, and F.M.~Peeters, 
Phys. Rev. B {\bf 70}, 144523 (2004).

\bibitem{wecorbino}
V.R.~Misko and F.M.~Peeters, 
Phys. Rev. B {\bf 74}, 174507 (2006).


\bibitem{book} 
J.B.~Ketterson and S.N.~Song, {\it Superconductivity}, 
(Cambridge University Press, 1999). 

\bibitem{md01}
F.~Nori, Science {\bf 278}, 1373 (1996);
C.~Reichhardt, J.~Groth, C.J.~Olson, S.~Field, and F.~Nori, Phys. Rev. B {\bf 52}, 
10~441 (1995); 
{\it ibid.} {\bf 53}, R8898 (1996);
{\bf 54}, 16~108 (1996);
{\bf 56}, 14~196 (1997). 

\bibitem{md0157}
C.~Reichhardt, C.J.~Olson, and F.~Nori, Phys. Rev. B {\bf 57}, 7937 (1998).

\bibitem{md02}
C.~Reichhardt, C.J.~Olson, and F.~Nori, Phys. Rev. Lett. {\bf 78}, 2648 (1997); 
Phys. Rev. B {\bf 58}, 6534 (1998). 

\bibitem{penrose}
V.~Misko, S.~Savel'ev, and F.~Nori, Phys. Rev. Lett. {\bf 95}, 177007 (2005); 
Phys. Rev. B {\bf 74}, 024522 (2006). 

\bibitem{bean-levingston}
C.P.~Bean and J.D.~Levingston, Phys. Rev. Lett. {\bf 12}, 14 (1964). 

\bibitem{tonomura-vvm}
K.~Harada, O.~Kamimura, H.~Kasai, T.~Matsuda, A.~Tonomura, and V.V.~Moshchalkov, 
Science {\bf 274}, 1167 (1996).

\bibitem{togawa}
Y.~Togawa, K.~Harada, T.~Akashi, H.~Kasai, T.~Matsuda, F.~Nori, A.~Maeda, 
and A.~Tonomura, Phys. Rev. Lett. {\bf 95}, 087002 (2005). 


\end{references}
\end{document}